\documentclass[a4paper,11pt]{article}

\usepackage{pos}
\usepackage{lipsum} 
\usepackage{siunitx}
\usepackage{graphicx}
\usepackage{subcaption}
\usepackage{todonotes}
\usepackage{bm}
\usepackage[inkscapelatex=false]{svg}
\usepackage{lineno}

\title{Using End-to-End Optimized Summary Statistics to Improve IceCube's Diffuse Galactic Fits}

\ShortTitle{Optimized Summary Statistics for Galactic Neutrino Measurements}

\author{The IceCube Collaboration \\{\normalsize \normalfont(a complete list of authors can be found at the end of the proceedings)}\\}

\emailAdd{oliver.janik@fau.de}
\emailAdd{christian.haack@fau.de}

\abstract{

Characterizing the astrophysical neutrino flux with the IceCube Neutrino Observatory traditionally relies on a binned forward folding likelihood approach. Insufficient Monte Carlo (MC) statistics in each bin limits the granularity and dimensionality of the binning scheme.
A neural network can be employed to optimize a summary statistic that serves as the input for data analysis, yielding the best possible outcomes. This end-to-end optimized summary statistic allows for the inclusion of more observables while maintaining adequate MC statistics per bin.
This work will detail the application of end-to-end optimized summary statistics in analyzing and characterizing the galactic neutrino flux, achieving improved resolution in the likelihood contours for selected signal parameters and models.

\vspace{4mm}

{\bfseries Corresponding authors:}
Oliver Janik$^{1*}$, 
Christian Haack$^{1}$\\
{$^{1}$ \itshape Erlangen Centre for Astroparticle Physics (ECAP), Friedrich-Alexander-Universität Erlangen-Nürnberg }\\[4mm]
$^*$ Presenter
}

\ConferenceLogo{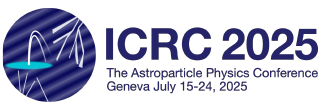}

\FullConference{39th International Cosmic Ray Conference (ICRC2025)\\
 15–24 July 2025\\
Geneva, Switzerland\\}

\begin{document}

\maketitle

\section{Introduction}

The IceCube Neutrino Observatory \cite{Aartsen:2016nxy} is a cubic-kilometer-scale neutrino telescope at the South Pole. It consists of 5160 digital optical modules (DOMs) embedded in the Antarctic glacier on 86 cables (strings) at depths between \qty{1450}{\m} and \qty{2450}{\m}. The DOMs house 10-inch photomultiplier tubes as well as digitization electronics. IceCube detects neutrinos by measuring the Cherenkov light emitted by charged secondaries produced in neutrino-matter interactions. 
IceCube has recently found evidence for neutrino emission from the Galactic Plane \cite{IceCubeGP, Fürst:2023, ESTESarxiv} in multiple detection channels. The next generation of analyses with increased livetime and a combination of multiple detection channels aims to increase the ability to differentiate between different galactic neutrino emission models as well as the precision of the measured model parameters. 

Current analysis methods require a set of high-quality observables (summary statistics) for each neutrino event. These observables are typically reconstructed properties such as the energy and direction. Using these observables, a likelihood is constructed. The probability distributions required for the likelihood are approximated using Monte Carlo (MC) simulations. These approximations generally suffer from the curse of dimensionality, which limits the amount of information that can be incorporated into the likelihood.

In this contribution, we present a method to increase the amount of information included in the likelihood and thereby provide an avenue to increase IceCube's capability to measure the galactic neutrino flux.

\section{Forward Folding Template Method}
One of the main analysis techniques used in IceCube is the forward folding template method. After the raw data from the detector have gone through several stages of data processing and filtering, a set of summary statistics is calculated for each event. These summary statistics summarize the properties of the event, such as the reconstructed energy and direction, the interaction vertex and estimators for the reconstruction quality. A subset of these quantities (typically the reconstructed energy and direction) is then used to construct a histogram. In order to perform a measurement, the expectation value $\lambda_j$ in each bin $j$ is calculated for the model of interest using MC simulations:
\begin{align}
    \lambda_j (\bm{\theta}) = \sum_i w_i(\bm{\theta}) \cdot I(\bm{\xi}_i) .
    \label{eq:bin_expectation}
\end{align}
where $w_i$ is the weight of each Monte Carlo event $i$, $\bm{\theta}$ are the model parameters and $I$ is the indicator function, which is $1$ if the event falls into bin $j$ given its summary statistics $\bm{\xi}_i$, and $0$ otherwise. The weights combine the instrument response $\nu_i$ with the differential flux prediction $\phi$ and the livetime $t$:
\begin{align}
w_i(\bm{\theta}) = \nu_i \left ( \bm{\theta}_\textbf{det} \right) \cdot t \cdot \phi \left (\bm{\theta}_\textbf{flux}\right ) .
\end{align}
This ensures that the expectation $\lambda$ can be recalculated efficiently for different model parameters $\bm{\theta}_\textbf{flux}$. The effect of systematic uncertainties (such as the detection efficiency of the DOMs, the optical properties of the Antarctic ice, and the atmospheric neutrino flux prediction) of the detector response ($\bm{\theta}_\textbf{det}$) can be parametrized using methods such as the \emph{Snowstorm} approach \cite{Snowstorm}. 

The signal parameters of interest are obtained via maximum-likelihood estimation. The total likelihood
$\mathcal{L}$ is computed by calculating the likelihood of observing $n_j$ events in bin $j$ given the expectation $\lambda_j$:
\begin{align}
    \mathcal{L}(\bm{\theta}) = \prod_j \mathcal{L}_j \left ( n_j \mid \lambda_j\left(\bm{\theta}\right) \right ) .
    \label{eq:llh} 
\end{align}
The per-bin likelihood $\mathcal{L}_j$ is either a conventional Poisson likelihood $\mathcal{L}_j = \frac{\exp \left (-\lambda_j \right )\cdot\lambda_j^{n_j}}{n_j!}$ or an effective Poisson likelihood \cite{say} accounting for the finite MC statistics in each bin.

In contrast to the unbinned analysis approach used in \cite{IceCubeGP}, systematic uncertainties can be efficiently incorporated into the test statistic by parameterizing their effects on the Monte-Carlo weights $w_i$ or the per-bin expectation $\lambda_j$. This allows for a robust estimation of the model parameters $\bm{\theta}$ at the cost of a reduced discovery potential introduced by the binning.

\section{End-to-End Optimized Summary Statistics}

\begin{figure}[h]
    \centering
    \includegraphics[width=1\linewidth]{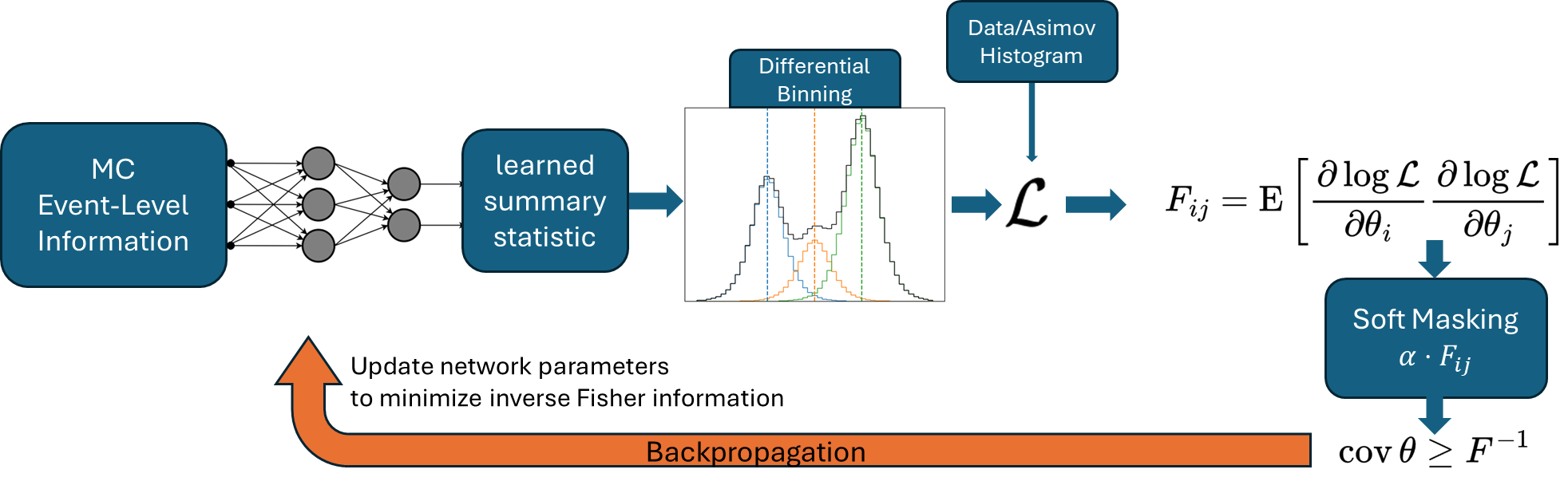}
    \caption{Workflow of the optimization process.}
    \label{fig:Pipeline}
\end{figure}

In the traditional binning approach, the number of bins scales exponentially with the number of dimensions. Since the binned forward folding approach requires sufficient MC statistics in each bin, this limits the number of variables used for binning. In addition, a rectangular binning scheme may not suit a measurement of a non-rectangular signal. Finding an optimal non-rectangular binning scheme in multiple dimensions is not trivial. 

Different machine learning-based algorithms have been proposed that automate this binning scheme optimization process, such as INFERNO \cite{Inferno} and neos \cite{neos}. This work is an extension to the neos framework, specifically for forward folding analyses.

To optimize the binning using gradients, a fully differentiable analysis is needed. The workflow for the optimization process is illustrated in Figure \ref{fig:Pipeline}. Event-level information is used as input for a neural network. The output of a network is the learned summary statistic $lss$. 
A differential binning method is used to calculate a histogram of $lss$. The counts of the histogram are then used to calculate the likelihood; see Equation \ref{eq:llh}. The likelihood is then multiplied by a masking factor $\alpha_i$, reducing the likelihood for bins with low MC statistics. Finally, the Cramér-Rao bound is utilized to estimate the variance of the signal parameters. From this estimated variance, a loss function is calculated. In the case of a single signal parameter, the variance is used directly. For multiple signal parameters, a weighted sum of the variances can be used. Using this loss function, the parameters of the neural network are optimized via gradient descent.

In the following, parts of the workflow are explained in detail.

\subsection{Differential Binning}

Binning data is a discrete process. A data point is sorted into a specific bin, incrementing its count by one. This process is not differentiable. INFERNO \cite{Inferno} lets a neural network output a vector where each entry corresponds to a count of a specific bin. In neos \cite{neos}, a neural network is used to compress the observables into a one-dimensional summary statistic. A kernel density estimate (KDE) approximates the distribution of the summary statistic. Finally, the cumulative density function of the KDE is used to calculate the bin-wise contributions for each event.

Our approach is similar to the neos approach. However, we replace the binned KDE with an approximation of a box function using the hyperbolic tangent. In a one-dimensional binning scheme, we can define the new indicator function $I$ for the $i$-th event as

\begin{equation}
    I_j\left(lss_i\right) = \mathcal{N}\left( \tanh\left(\frac{lss_i-b_j}{s}\right) \tanh\left(\frac{b_{j+1}-lss_i}{s}\right) +1\right) .
\end{equation}
Here, $b_j$ is the left edge of the $j$-th bin, $\mathcal{N}$ is a normalization factor that chosen such that $\sum_in_i=1$, $s$ is the slope parameter where $\lim_{s\to0}$ will recover the box function.

For the case of more than one dimension for the optimized binning, the identity function $I_{i_d}$ is evaluated for each dimension individually and combined using the outer product over all dimensions $d$

\begin{equation}
    I_{i_1,...,i_D}\left(lss_1,...,lss_D\right) = \bigotimes_{d=1}^{D} I_{i_d}\left(lss_d\right) .
\end{equation}

\begin{figure}
    \centering
    \includegraphics[width=0.5\linewidth]{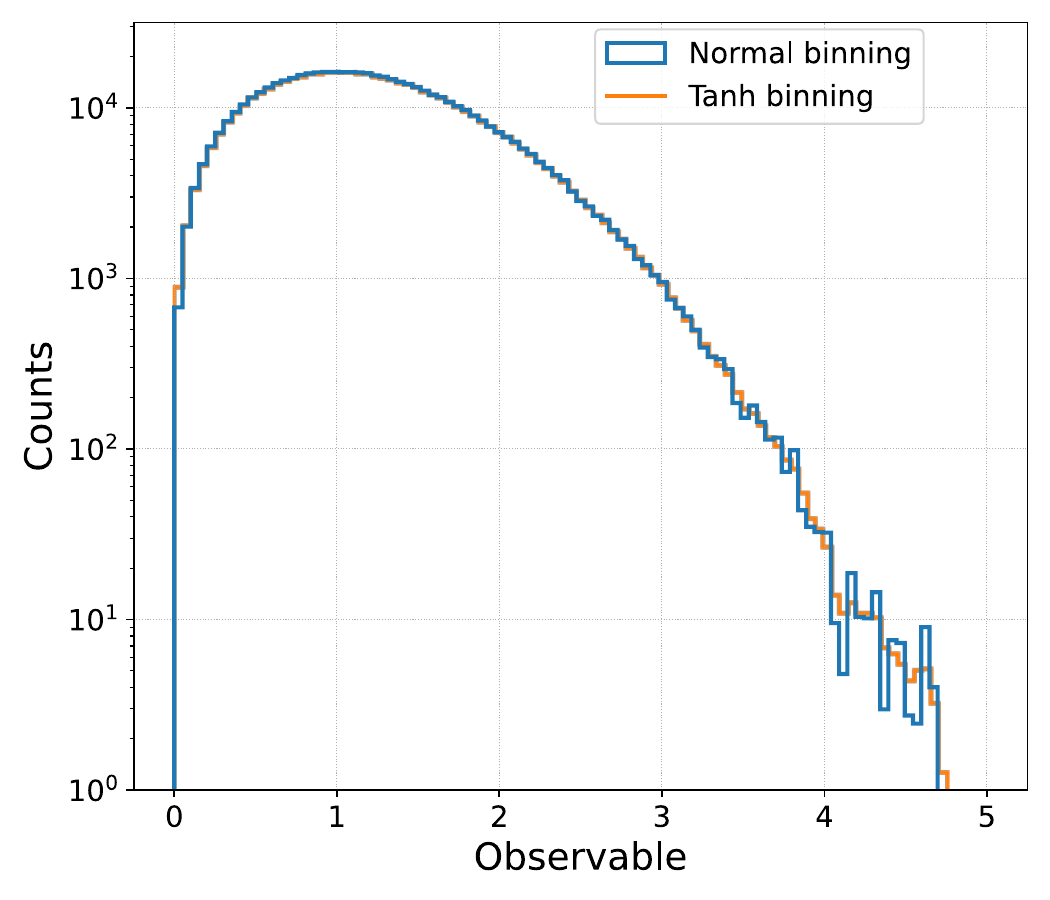}
    \caption{Comparison between a normal histogram and one created using the differential tanh binning. The histogram includes $\num{1e6}$ randomly generated data points with randomly generated weights.}
    \label{fig:enter-label}
\end{figure}

\subsection{Fisher Information}

The variance of a signal parameter can be estimated by doing a likelihood scan, based on Wilks' theorem. This requires performing multiple likelihood fits for a set of fixed values for the signal parameter. For the optimization process, where multiple evaluations of the variance are needed, this process is not feasible. Instead, we calculate the Fisher information matrix:

\begin{equation}
    F_{ij} = \mathrm{E}\left[\frac{\partial \log\mathcal{L}}{\partial\theta_i}\frac{\partial \log\mathcal{L}}{\partial\theta_j}\right] .
\end{equation}
We can now utilize the Cramér-Rao bound, stating that the inverse of the Fisher information is a lower limit of the covariance of the signal parameters:

\begin{align}
    \mathrm{cov}\bm{\theta}\geq F^{-1} .
\end{align}
Using the Poisson likelihood, the Fisher information can be analytically solved as

\begin{align}
    F_{ij} = \frac{1}{\lambda} \frac{\partial\lambda}{\partial\theta_i}\frac{\partial\lambda}{\partial\theta_j} .
\end{align}

\subsection{Soft Masking}

To account for finite MC statistics in each bin, an effective likelihood can be used \cite{say}. Here, in addition to the expectation $\lambda_j$ (see Equation \ref{eq:bin_expectation}), the error of the expectation is calculated as

\begin{align}
    \sigma_k^2 (\bm{\theta}) = \sum_i w_i(\bm{\theta})^2 \cdot I(\bm{\xi}_i) \, .
\end{align}
Instead of implementing this effective likelihood into the optimization pipeline, a penalty is added to punish the network for producing bins with low MC statistics. Therefore, the relative uncertainty of a bin is defined as $\delta_k = \frac{\sigma_k}{\lambda_k}$. To reduce $\delta_k$, an additional loss can be introduced that increases with a higher relative uncertainty. However, this led in most cases to an unstable training routine. 

Instead, a masking routine is introduced. If a bin has insufficient MC statistics, it gets masked out and will not contribute to the Fisher information. To make this masking differentiable, a soft cut-off is used, utilizing a sigmoid function. The masking factor is given by:

\begin{align}
    \alpha_k = 1 - \frac{1}{1+\exp\left(\frac{\delta_k-t}{z}\right)} .
\end{align}
Here, $t$ is a threshold value, and $z$ is the sharpness of the sigmoid's edge. The new likelihood value of a bin is now given by $\alpha_k\cdot F_{ij,k}$. For $\delta_j\gg t$, $\alpha_j$ becomes zero, and the bin will not contribute to the total likelihood calculation. 

\section{Optimized Binning for a Galactic Plane Measurement}

The End-to-End optimization is now applied to a measurement of the neutrino flux from the Galactic plane. The Galactic plane is modeled using the CRINGE template \cite{Schwefer_2023}. The template is scaled by the normalization factor $\mathrm{\Psi_{CRINGE}}$, which will be our signal parameter. Nuisance parameters are included for a single powerlaw spectrum describing the isotropic astrophysical neutrino flux, as well as multiple nuisance parameters for modeling the atmospheric neutrino flux. The binning will be optimized to reduce $\mathrm{Var\left(\Psi_{CRINGE}\right)}$.

We use the IceCube Northern Tracks sample \cite{NT2022}, consisting of upgoing muon tracks. This sample is a high-purity neutrino sample, using the Earth as a shield for atmospheric muons. These neutrino-induced tracks allow for good pointing to their source. However, since IceCube is located on the southern hemisphere, this selection does not include neutrinos from the galactic center, which is located in the southern sky. 

\begin{figure}[h]
    \centering
    \begin{subfigure}[b]{0.48\linewidth}
        \centering
        \includegraphics[width=\linewidth]{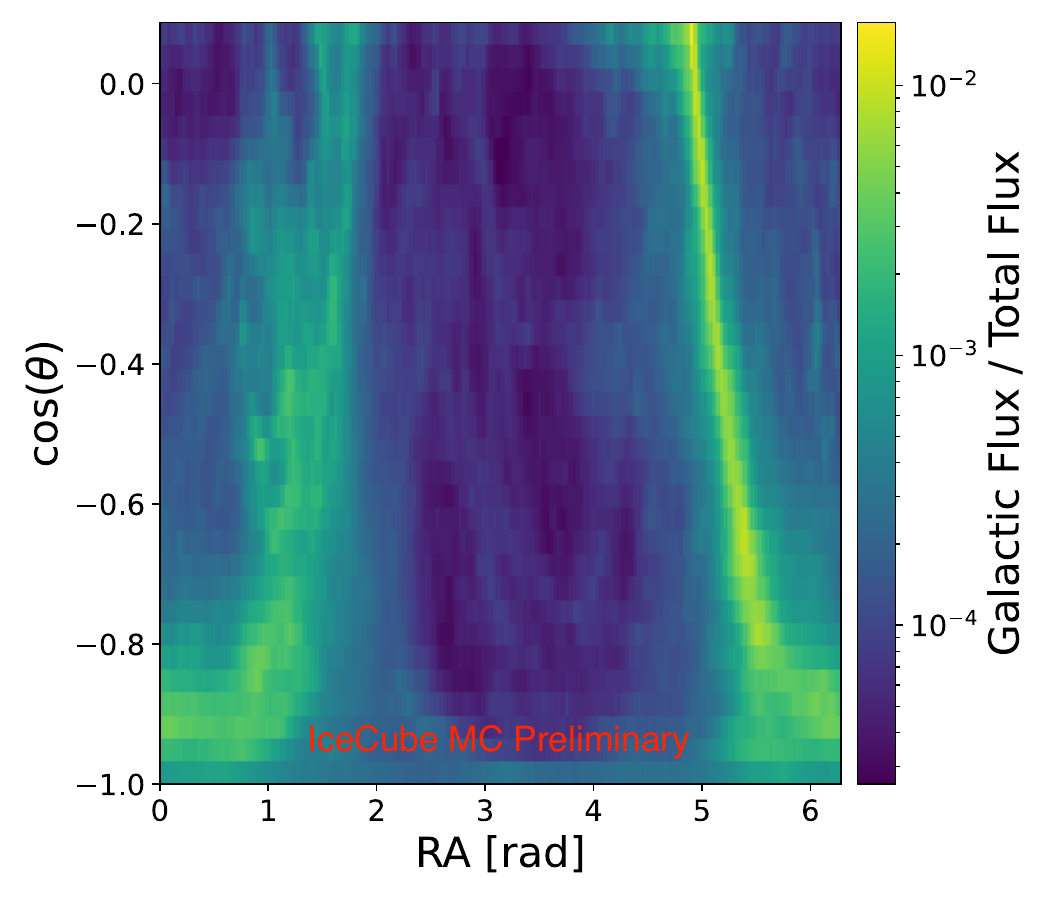}
        \caption*{(a)}
        \label{fig:galactic-standard}
    \end{subfigure}
    \hfill
    \begin{subfigure}[b]{0.48\linewidth}
        \centering
        \includegraphics[width=\linewidth]{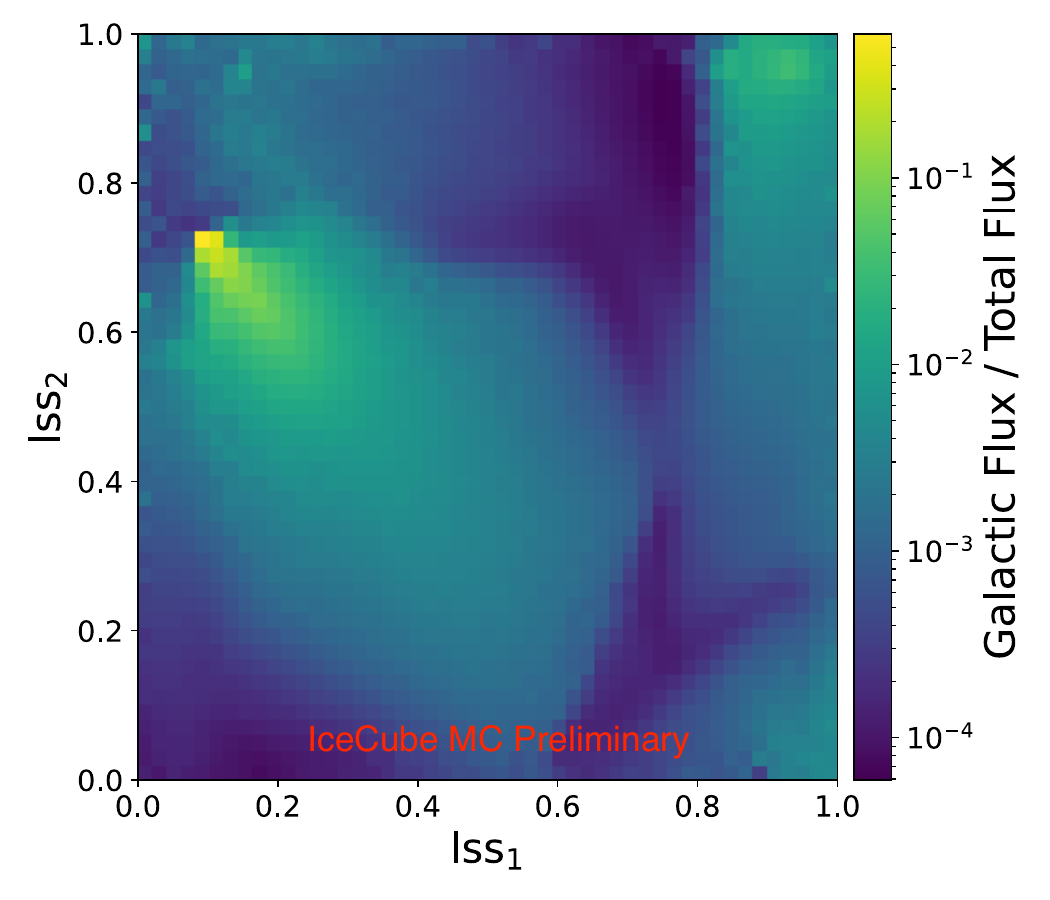}
        \caption*{(b)}
        \label{fig:galactic-optimized}
    \end{subfigure}
    \caption{Comparison between the galactic flux in (a) a two-dimensional projection of the standard binning and (b) the optimized binning. The standard binning shows the physical shape of the Galactic plane. As IceCube is located at the South Pole the galactic center can not be seen. In the optimized binning the Galactic plane appears as an undefined shape in the upper left corner.}
    \label{fig:combined-galactic}
\end{figure}

The performance of the optimized binning will be compared with the standard three-dimensional binning used in similar analyses \cite{Fürst:2023}. The standard binning includes 45 bins in reconstructed energy, 33 bins in reconstructed cosine zenith, and 180 bins in reconstructed right ascension. In total, the binning scheme consists of \num{267300} bins. A two-dimensional projection of the standard binning showing the ratio between the CRINGE prediction of the galactic flux and the total expected flux is shown in the left panel of Figure \ref{fig:combined-galactic}. 

The three variables used for the standard binning serve as input to the neural network. In addition, variables are added that describe the quality of the reconstruction. These variables were previously used to distinguish upgoing from downgoing muon tracks in the detector \cite{Rädel:2017} and include, among others, an estimate of the angular uncertainty for each event.

The network architecture is a fully connected model with 5 hidden layers with 64 to 512 nodes each. The number of output nodes corresponds to the dimensionality of the optimized binning. A comparison shows that a two-dimensional binning minimizes the loss the best in this particular case. Therefore, two output nodes are used in the network. The output is then scaled between 0 and 1. The two summary statistics $lss_1$ and $lss_2$ are each binned using 40 bins, making 1600 bins in total. To guarantee enough MC statistics, softmasking was used with a threshold of 5\%.

The ratio between the predicted galactic flux and the predicted total flux in the optimized binning scheme can be seen in Figure \ref{fig:combined-galactic}. The Galactic plane appears as a concentration in the upper left corner around $lss_1=0.7$ and $lss_2=0.1$. An initial investigation of this region has revealed that mostly high-energy, well-reconstructed events are placed in that region. 

For comparison, an Asimov scan is performed \cite{Cowan2011}. This allows us to estimate the performance of the binning using MC simulation only. As our baseline model assumption, we use the baseline CRINGE flux prediction.

\begin{figure}[h]
    \centering
    \begin{subfigure}[b]{0.48\linewidth}
        \centering
        \includegraphics[width=\linewidth]{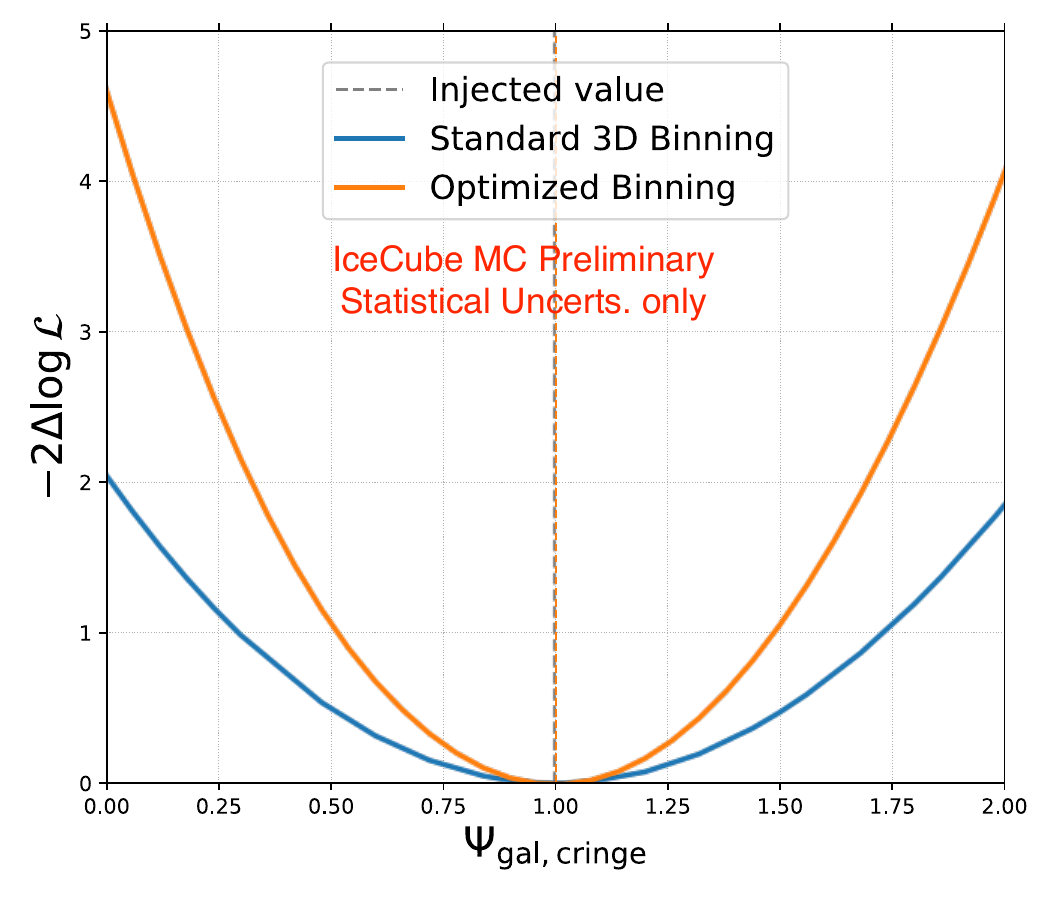}
        \caption*{(a)}
        \label{fig:scan-1}
    \end{subfigure}
    \hfill
    \begin{subfigure}[b]{0.48\linewidth}
        \centering
        \includegraphics[width=\linewidth]{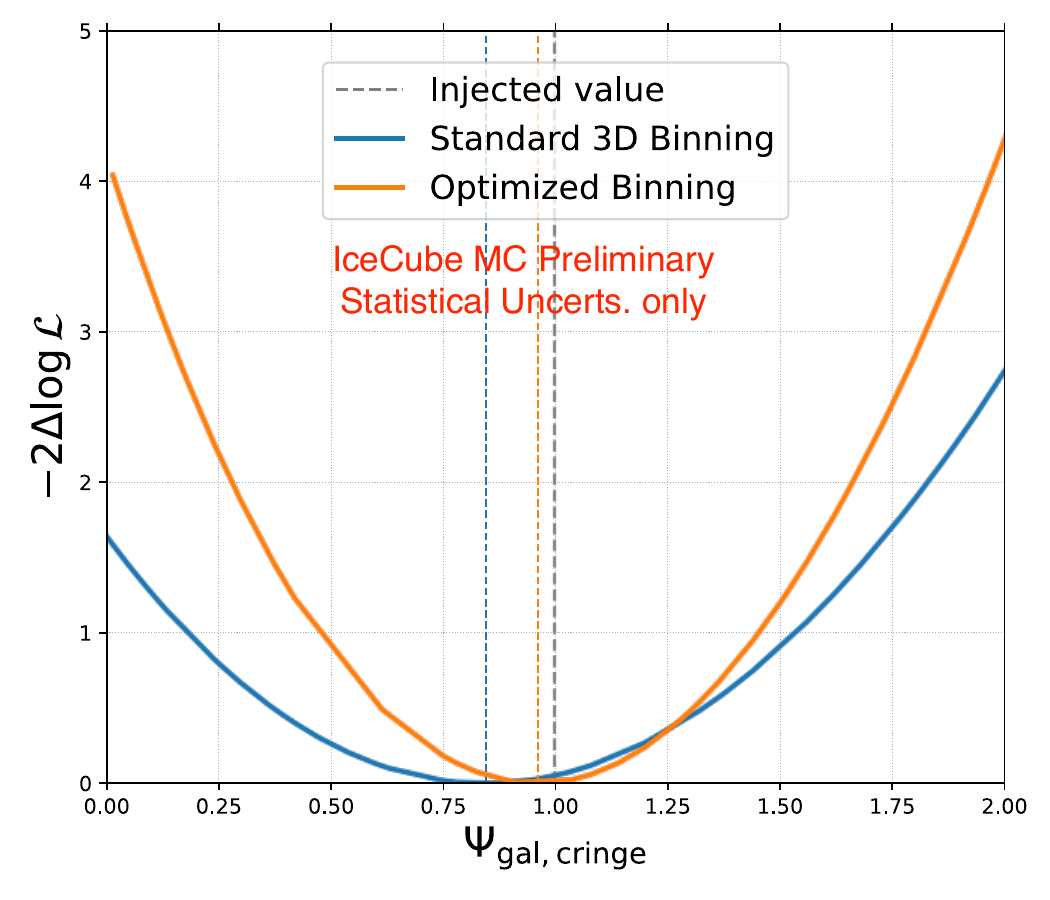}
        \caption*{(b)}
        \label{fig:scan-2}
    \end{subfigure}
    \caption{Profile likelihood scan for (a) the Poisson likelihood and (b) and effective likelihood. In both cases the optimized binning achieves a smaller uncertainty region. In case of the effective likelihood the optimized binning also achieves a smaller bias.}
    \label{fig:scan}
\end{figure}

The resulting Poisson likelihood scan is shown in the left panel of Figure \ref{fig:scan}. Both the standard and the optimized binning scheme recover the injected values of $\Psi_\mathrm{cringe}$, as expected from an Asimov likelihood scan.
For the standard binning $\Psi_{cringe}=\num{0}$, corresponding to no galactic neutrinos, can be excluded with $-\Delta\log\mathcal{L}=2$, which is about $1.4\sigma$. The optimized binning can exclude no galactic neutrinos with $-\Delta\log\mathcal{L}=4.5$ or $2.1\sigma$. The $\num{1}\sigma$ interval of the estimator is improved by $\sim\num{33}\%$ for the optimized binning.

The scan using the effective likelihood (see Figure \ref{fig:scan}, right panel) shows a reduced bias compared to the standard binning while maintaining a similar improvement on the  $\num{1}\sigma$ interval. The bias is a consequence of the finite MC statistics. 

\section{Conclusion}

We have shown that utilizing an optimized summary statistic improves the measurement of the galactic neutrino flux. This improvement is achieved by introducing variables that quantify the quality of the neutrino property reconstruction to a summary statistic optimized by a neural network. This summary statistic can be used to build a histogram that is used for a forward folding likelihood analysis. Here, this binning achieves a better performance than the previously used binning scheme while using far fewer bins. This has the additional advantage of reducing uncertainties introduced by finite MC statistics for computing the per-bin expectations. This method also finds application in other IceCube analyses, such as a measurement of the astrophysical neutrino flavor composition \cite{vaneeden:2025icrc}.

Currently, such analyses are based on high-level data selections. Fixed cuts on reconstructed parameters are used. In the future, the method presented here may be used to optimize these cuts by introducing the corresponding parameters to the input of the optimization routine. 

\bibliographystyle{ICRC}
\bibliography{references}

%

\clearpage

\section*{Full Author List: IceCube Collaboration}

\scriptsize
\noindent
R. Abbasi$^{16}$,
M. Ackermann$^{63}$,
J. Adams$^{17}$,
S. K. Agarwalla$^{39,\: {\rm a}}$,
J. A. Aguilar$^{10}$,
M. Ahlers$^{21}$,
J.M. Alameddine$^{22}$,
S. Ali$^{35}$,
N. M. Amin$^{43}$,
K. Andeen$^{41}$
% ...
\\
\\
% ...
$^{10}$ Universit{\'e} Libre de Bruxelles, Science Faculty CP230, B-1050 Brussels, Belgium \\
% ...
$^{16}$ Department of Physics, Loyola University Chicago, Chicago, IL 60660, USA \\
$^{17}$ Dept. of Physics and Astronomy, University of Canterbury, Private Bag 4800, Christchurch, New Zealand \\
% ...
$^{21}$ Niels Bohr Institute, University of Copenhagen, DK-2100 Copenhagen, Denmark \\
$^{22}$ Dept. of Physics, TU Dortmund University, D-44221 Dortmund, Germany \\
% ...
$^{35}$ Dept. of Physics and Astronomy, University of Kansas, Lawrence, KS 66045, USA \\
% ...
$^{39}$ Dept. of Physics and Wisconsin IceCube Particle Astrophysics Center, University of Wisconsin{\textemdash}Madison, Madison, WI 53706, USA \\
% ...
$^{41}$ Department of Physics, Marquette University, Milwaukee, WI 53201, USA \\
% ...
$^{43}$ Bartol Research Institute and Dept. of Physics and Astronomy, University of Delaware, Newark, DE 19716, USA \\
% ...
$^{63}$ Deutsches Elektronen-Synchrotron DESY, Platanenallee 6, D-15738 Zeuthen, Germany \\
$^{\rm a}$ also at Institute of Physics, Sachivalaya Marg, Sainik School Post, Bhubaneswar 751005, India \\
% ...

\subsection*{Acknowledgments}

\noindent
The authors gratefully acknowledge the support from the following agencies and institutions:
USA {\textendash} U.S. National Science Foundation-Office of Polar Programs,
U.S. National Science Foundation-Physics Division,
U.S. National Science Foundation-EPSCoR,
U.S. National Science Foundation-Office of Advanced Cyberinfrastructure,
Wisconsin Alumni Research Foundation,
Center for High Throughput Computing (CHTC) at the University of Wisconsin{\textendash}Madison,
Open Science Grid (OSG),
Partnership to Advance Throughput Computing (PATh),
Advanced Cyberinfrastructure Coordination Ecosystem: Services {\&} Support (ACCESS),
Frontera and Ranch computing project at the Texas Advanced Computing Center,
U.S. Department of Energy-National Energy Research Scientific Computing Center,
Particle astrophysics research computing center at the University of Maryland,
Institute for Cyber-Enabled Research at Michigan State University,
Astroparticle physics computational facility at Marquette University,
NVIDIA Corporation,
and Google Cloud Platform;
Belgium {\textendash} Funds for Scientific Research (FRS-FNRS and FWO),
FWO Odysseus and Big Science programmes,
and Belgian Federal Science Policy Office (Belspo);
Germany {\textendash} Bundesministerium f{\"u}r Forschung, Technologie und Raumfahrt (BMFTR),
Deutsche Forschungsgemeinschaft (DFG),
Helmholtz Alliance for Astroparticle Physics (HAP),
Initiative and Networking Fund of the Helmholtz Association,
Deutsches Elektronen Synchrotron (DESY),
and High Performance Computing cluster of the RWTH Aachen;
Sweden {\textendash} Swedish Research Council,
Swedish Polar Research Secretariat,
Swedish National Infrastructure for Computing (SNIC),
and Knut and Alice Wallenberg Foundation;
European Union {\textendash} EGI Advanced Computing for research;
Australia {\textendash} Australian Research Council;
Canada {\textendash} Natural Sciences and Engineering Research Council of Canada,
Calcul Qu{\'e}bec, Compute Ontario, Canada Foundation for Innovation, WestGrid, and Digital Research Alliance of Canada;
Denmark {\textendash} Villum Fonden, Carlsberg Foundation, and European Commission;
New Zealand {\textendash} Marsden Fund;
Japan {\textendash} Japan Society for Promotion of Science (JSPS)
and Institute for Global Prominent Research (IGPR) of Chiba University;
Korea {\textendash} National Research Foundation of Korea (NRF);
Switzerland {\textendash} Swiss National Science Foundation (SNSF).

\end{document}